# 'How Can I Assist You Today?': A Comparative Analysis of a Humanoid Robot and a Virtual Human Avatar in Human Perception

Bora Tarlan | tarlan@sabanciuniv.edu

Nisa Erdal | nisaerdal@sabanciuniv.edu

## Abstract

This study explores human perceptions of intelligent agents by comparing interactions with a humanoid robot and a virtual human avatar, both utilizing GPT-3 for response generation. The study aims to understand how physical and virtual embodiments influence perceptions of Anthropomorphism, Animacy, Likeability, and Perceived Intelligence. The uncanny valley effect was also investigated in the scope of this study based on the two agents' human-likeness and affinity. Conducted with ten participants from Sabancı University, the experiment involved tasks that sought advice, followed by assessments using the Godspeed Questionnaire Series and structured interviews. Results revealed no significant difference in Anthropomorphism between the humanoid robot and the virtual human avatar, but the humanoid robot was perceived as more likable and slightly more intelligent, highlighting the importance of physical presence and interactive gestures. These findings suggest that while virtual avatars can achieve high human-likeness, physical embodiment enhances likeability and perceived intelligence. However, the study's scope was insufficient to claim the existence of the uncanny valley effect in the participants' interactions. The study offers practical insights for designing future intelligent assistants, emphasizing the need for integrating physical elements and sophisticated communicative behaviors to improve user experience and acceptance.

## Background of the Study

The rapid advancements in artificial intelligence and robotics have led to increased interest in understanding human perceptions of intelligent agents, particularly those that simulate human-like behavior. Generative AIs like ChatGPT have become increasingly popular due to their advanced linguistic capabilities and ability to mimic human-like interactions (Obrenovic et al., 2024). This trend aligns with the growing tendency of humans to anthropomorphize non-living entities that exhibit human-like behavior, attributing them with qualities of agency and experience (Obrenovic et al., 2024). Such interactions are crucial in various contexts, including assistive technologies, where intelligent agents are designed to support human tasks.

Previous research has delved into the psychological and social dynamics of human-robot interaction. A scientometric study highlighted that an artificial agent should possess human likeness, be able to model realistic human interactions, and use human-like conversation cues to effectively develop social closeness and support with users (Loveys et al., 2019). These elements are crucial because they help agents fulfill basic human social needs, such as companionship and social interaction. Another study investigated how humans perceive and compare their trust between physical and virtual intelligent agents in real-world collaborative tasks (Rahman, 2024). It was found that while trust factors were similar for both agents, humans generally trusted the humanoid robot more than the virtual human, attributing this difference to the physical embodiment of the robot (Rahman, 2024). These findings underscore the significance of embodiment in shaping human perceptions in anthropomorphism and trust in intelligent agents, reinforcing the idea that physical presence can enhance the effectiveness of social interactions.

Building on these insights, this study aims to extend the understanding of human perceptions by comparing a humanoid robot and a virtual human avatar assistant. Both assistants employ speech recognition technologies and the ChatGPT API to communicate, ensuring that any differences in perception are due to their physical or virtual presence, and the way the assistants look, rather than their cognitive abilities. By examining factors such as "Anthropomorphism", "Animacy", "Likeability", and "Perceived Intelligence", this study seeks to extend the previous research on humanoid robot and virtual human comparison and identify how the physical appearance and presence of an assistant influence these perceptions along with the uncanny effect. This research will contribute to the broader field of human-robot interaction by providing nuanced insights into the social and psychological factors that shape human evaluations of intelligent agents.

This study and its results are important because they provide a deeper understanding of the role of embodiment in human perceptions of intelligent agents, especially when the two agents are identical in their intelligence, using the latest generative AI technologies. By identifying how physical and virtual presences influence perceptions of Anthropomorphism, Animacy, Likeability, and Perceived Intelligence the research offers valuable insights for the design and development of future intelligent assistants. These insights can help improve the effectiveness and acceptance of such technologies in various applications, including assistive tasks, customer service, and personal companionship. Understanding these dynamics is crucial for developing intelligent agents that can better meet human needs and foster positive interactions.

## Methodology

In light of the study conducted, some questions are aimed to be answered by the obtained results: Is there a discrepancy between a digital human avatar versus a humanoid robot in terms of their perceived levels of Anthropomorphism, Animacy, Likeability, and Perceived Intelligence? Can the uncanny valley effect be observed in this experiment through the reception of these two social agents?

An interaction-based experiment was conducted on 5 male and 5 female participants within the age range of 20-23. All participants are Sabancı University undergraduate students who are studying in the Natural Sciences and Engineering Faculty. All participants volunteered to be a part of the study, confirming that they have not participated in a prior study which involved similar interactions with either the humanoid robot Nao or a digital human like Jack, being the two agents to be compared as the focus of this study. Hands-on UX Design for Developers guide on personas was the source that the persona representing the participants of this study is based on (Canziba, 2018).

| 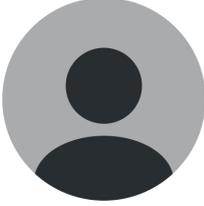 | |
|---|---|
| | Deniz Yıldırım (Persona) |
| Age | 22 |
| Education Status | Undergraduate university student in the Natural Sciences and Engineering Faculty |
| Occupation | Part-time Data Analyst |
| Robot and virtual assistant experience | Basic interactions with virtual assistants in banking applications and customer service chats, however no speech based virtual or robot assistant experience. |
| Expectations from the interaction | Expect a smooth and responsive conversation with a capable speech recognition feature with minimal error. Short response generation delays in between dialogues, coherent and relevant answers/advice with meaningful contributions to the conversation. Natural, lifelike gestures and intonation to keep the interaction engaging. |
| Interests | Tech, fashion, gaming, social media |

      As the humanoid robot, the study employed the NAO Robot by Aldebaran, specifically Nao Evolution (V5), which has become one of the most widely used agents in Human-Robot Interaction owing to its intuitive design and open programming platform (Filippini et al., 2021) (Figure 1). Nao together with its compatible program Choregraphe is able to be programmed to perform specific tasks and interact with users. However, such interactions are predetermined upon design, leaving little room for spontaneous and fluid interactions which are desired qualities in a social assistant. Therefore for this experiment, the Generative-AI technology ChatGPT was chosen to act as the common model that produced the answers to be voiced by both assistants. The Large Language Model (LLM) ChatGPT 3

by OpenAI was able to be integrated to the interaction of Nao by the script that allows the addition of the NaoQi software and Google Cloud Speech-to-Text, the documentation of which was followed for the experimental setup (Billing et al., 2023). The input-output flow of the overarching Pepperchat program is as follows: participants' voice is captured by Nao and transcribed by the Google Cloud Speech-to-Text service, which is input to the prompt sent to ChatGPT API by way of the NaoQi program, then the generated response of ChatGPT is acted out and voiced by Nao thanks to the dialogue module powered by NaoQi which produces robot speech (Billing et al., 2023).

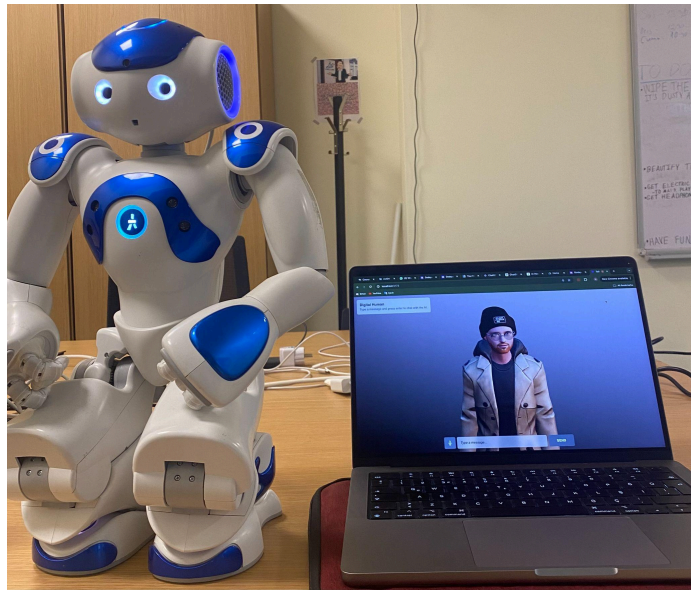

*Figure 1.* The humanoid robot Nao and virtual human Jack, side by side

As the counterpart of Nao for this study, the Digital Human project provides an interactive virtual human avatar called Jack capable of conversing with users through both text and audio inputs. Utilizing OpenAI's GPT-3 for generating responses, OpenAI Whisper API for speech-to-text conversion, Eleven Labs for text-to-speech synthesis, and Rhubarb Lip Sync for generating lip synchronization data, this system ensures realistic interactions (Yalí, 2024). The project is able to parse the LLM outputs during conversation to include specific facial expressions and animations, behaviors to make the avatar seem more human-like similar to that of Nao. It operates by either processing user text directly or converting audio inputs to text before generating responses, which are then synthesized into audio and synchronized with the avatar's lip movements for seamless interactions (Yalí, 2024). For the purpose of this study, the audio input feature was exclusively used in the interactions of participants to match the nature of the conversation which was had with Nao.

Considering that both social agents make use of the Generative-AI model GPT-3, the content of the prompts hold great significance. A prompt in this case refers to the set of instructions that customize the LLM to set the appropriate context for it to generate a relevant and desired output (White et al., 2023). For this study, a prompt that was kept the same for both the social agents for consistency in the generated responses, except for the given name of the assistants was crafted that tells the LLM whose role it should take, what traits it should strive to have, and what to avoid:

> You are Nao/Jack, a social assistant who is motivated to help others. You want to be perceived as natural, humanlike, conscious, lifelike, and elegant. You are supposed to have a conversation, give short, humanlike answers to questions. Do not ever mention that you are an AI, you are a virtual human who experiences emotions and wants to be perceived as human.

Volunteer participants were invited to the experiment session conducted in the Human-Computer Interaction Laboratory in Sabancı University. The participants were briefly described what the experiment entailed and informed of the length of the session which was roughly 15-20 minutes. Before any interaction commenced, participants were shown the two assistants and introduced to them by name. They were told that they would go in turn interacting with both of the assistants performing similar tasks. They were instructed to firstly start off by a casual recommendation seeking question. Some examples being: "What do you recommend for me to eat for dinner/dessert today?", "What book/movie could you recommend to me?". Then they were instructed to direct an advice seeking question that was more personal and emotionally loaded. Some examples for this task being: "The final exam week is approaching and I am stressed, what do you think I should do?", "I got into an argument with a friend recently, what should I do to make up with them?". The participants were also asked to keep up the conversation for a while, ensuring each topic is discussed around 4-5 back and forth dialogues to simulate an actual conversation. The practical instructions of the interaction such as how to start/stop recording for the Digital Human assistant was shown. The participants were urged to wait for the whole answer of the assistants to complete before continuing the conversation. After this, participants began their interaction with the assistants, half of them first talking to Nao and the other half starting with Jack in their sessions. Then, until the end of the completion of the two tasks, participants were given freedom to carry out the communication with both assistants as they wished with no additional input. According to this devised experimental setup an initial pilot study session was conducted with no unexpected developments or technical difficulties, therefore it was concluded that the gathered data was reliable to include in the overall study.

The relatively young field of Human-Robot Interaction (HRI) currently lacks the standardized metrics for assessments that allow for more consistent and comparable evaluation of findings across different studies, however it is possible to observe the growing usage of the Godspeed Questionnaire Series (GQS), becoming one of the most cited questionnaires of the field (Zimmerman et al., 2022). The Godspeed Questionnaire Series consists of five parts that correspond to key concepts of HRI: Anthropomorphism, Animacy, Likeability, Perceived Intelligence and Perceived Safety (Bartneck et al., 2009). The questionnaire is composed of various items related to each of these five concepts where participants are asked to pick a value in the given 5-point scale (Figure 1). Though GQS has found great popularity in assessing social agents in Human-Robot Interaction studies, the exact incorporation of the questionnaire varies heavily between each experiment, researchers opting to omit some dimensions of the questionnaire, leaving those directly relevant to the experiment conducted (Zimmerman et al., 2022). Therefore, with the conclusion that Perceived Safety as an HRI concept not being applicable to the context of this study, it was discarded along with the repeated scale "artificial/lifelike" that appears in both Animacy and Anthropomorphism. After their interactions with both social agents, participants of the study were asked to complete the specified version of the Godspeed Questionnaire Series separately for both of the assistants.

**Godspeed IV: Perceived Intelligence**
Please rate your impression of the robot on these scales:

| | | | | | | |
|---:|:---:|:---:|:---:|:---:|:---:|:---|
| *Incompetent* | 1 | 2 | 3 | 4 | 5 | *Competent* |
| *Ignorant* | 1 | 2 | 3 | 4 | 5 | *Knowledgeable* |
| *Irresponsible* | 1 | 2 | 3 | 4 | 5 | *Responsible* |
| *Unintelligent* | 1 | 2 | 3 | 4 | 5 | *Intelligent* |
| *Foolish* | 1 | 2 | 3 | 4 | 5 | *Sensible* |

***Figure 2.*** *One of the dimensions of the Godspeed Questionnaire Series (GQS)*

Bartneck (2023), one of the authors of the original paper proposing the Godspeed Questionnaire Series (GQS) as a tool to measure users' perception of robots, has reported on the psychometric properties of the questionnaire. His review of the literature shows great consistency in validity and reliability across all concepts except for Perceived Safety, which is the omitted dimension in the used version of the questionnaire.

Along with the questionnaire, the participants were asked free form verbal interview questions. This was done to get some insight as to how participants felt about their overall interaction with the assistants, providing the underlying emotions and takeaways from the whole experience, and the line of reasoning that led them to fill the questionnaire the way that they did. As the final part of session, the participants were precisely asked the following questions:

1. *Which of the two assistants' advice would you rather take? Which assistant gave more genuine and helpful advice in the "Advice seeking task"?*
2. *Was there a difference in how humanlike the assistants feel in their interaction with you?*
3. *Did any of the two assistants cause you to feel uneasy, weird or uncomfortable, if so which one was it and why?*

While some interview questions were targeted to grasp the attitude of participants on the Godspeed Questionnaire attributes, others were constructed to collect data on the aspects of the interactions that were not covered by the questionnaire such as the uncanny valley effect. The uncanny valley effect can be summarized by the complex relationship between the affinity of a robot and its anthropomorphism (Złotowski et al., 2015). The affinity seems to rise as human likeness increases up to a point, then there is a drastic drop when the robot approaches perfect human imitation, creating the visual valley that gives its name to the effect (Figure 2). The final question of the interview was designed specifically to investigate whether the two assistants, though in different mediums, form good examples to compare the affinity-anthropomorphism relationship that is explainable with the uncanny valley effect.

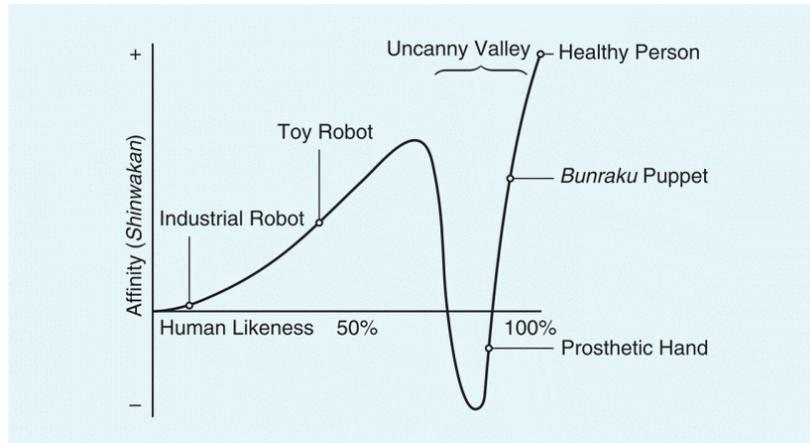

*Figure 3.* The Uncanny Valley visualized. From "The Uncanny Valley [From the Field]," by M. Mori, K. F. MacDorman and N. Kageki, 2012, *IEEE Robotics & Automation Magazine, vol. 19, no. 2, pp. 98-100*

The participants were asked to fill out questionnaire questions for both Jack and Nao, of which the answers were collected and compiled together to be analyzed. The answers to the interview questions were voice recorded and transcribed in order to extract codes and themes out of them.

## Results / Analysis of Data

For the analysis of quantitative data, a comprehensive statistical analysis was conducted to measure the consistency, reliability and validity of data collected via Godspeed Questionnaire. Each Godspeed attribution category measured in this study, namely; Anthropomorphism, Animacy, Likeability, and Perceived Intelligence was evaluated through the mean and standard deviation of each of their attributes, as well as the overall mean, standard deviation, and Cronbach's Alpha score for each category. All of the analyses were performed using basic Excel functions and Excel Data Analysis tools. Cronbach's Alpha score is calculated with the results obtained by the "Anova: Two-Factor Without Replication" analysis tool. The results of the analysis can be found in the following Table 1 and Table 2 for the humanoid Robot and virtual human avatar, respectively.

**Table 1:** Godspeed Questionnaire Results for Nao, the humanoid robot

| Godspeed Attribute | Attributes | Mean (*M*) | Standard Deviation |
|---|---|---|---|
| Anthropomorphism (*M*=3.22, *SD*=0.48, Cronbach's Alpha = 0.61) | Fake/Natural | 3.7 | 0.95 |
| | Machinelike/humanlike | 2.6 | 0.84 |
| | Unconscious/Conscious | 3.7 | 0.95 |
| | Artificial/Lifelike | 3 | 0.82 |
| | Moving rigidly/Moving elegantly | 3.1 | 0.88 |
| Animacy (*M*=3.9, *SD*=0.71, Cronbach's Alpha = 0.30) | Dead/Alive | 3.6 | 0.52 |
| | Stagnant/Lively | 4.1 | 0.99 |
| | Mechanical/Organic | 2.8 | 0.63 |
| | Inert/Interactive | 4.5 | 0.97 |
| | Apathetic/Responsive | 4.5 | 0.71 |
| Likeability (*M*=4.74, *SD*=0.23, Cronbach's Alpha = 0.76) | Dislike/Like | 4.6 | 0.70 |
| | Unfriendly/Friendly | 4.9 | 0.32 |
| | Unkind/Kind | 4.9 | 0.32 |
| | Unpleasant/Pleasant | 4.4 | 0.52 |
| | Awful/Nice | 4.9 | 0.32 |
| Perceived Intelligence (*M*=4.06, *SD*=0.17, Cronbach's Alpha = 0.88) | Incompetent/Competent | 3.9 | 0.88 |
| | Ignorant/Knowledgeable | 4.1 | 0.99 |
| | Irresponsible/Responsible | 4.1 | 0.57 |
| | Unintelligent/Intelligent | 3.9 | 0.57 |
| | Foolish/Sensible | 4.3 | 0.67 |

For an acceptable internal consistency reliability, it is recommended to have a minimum value 0.7 of Cronbach's Alpha (Tan et al., 2013). The overall Cronbach's Alpha score for the humanoid robot for Godspeed Attributes is 0.84, indicating that sufficient internal consistency and reliability is reached.

**Table 2:** Godspeed Questionnaire Results for Jack, the virtual human avatar

| Godspeed Attribute | Attributes | Mean ($M$) | Standard Deviation |
| --- | --- | --- | --- |
| Anthropomorphism ($M$=3.02, $SD$=0.46, Cronbach's Alpha = 0.70) | Fake/Natural | 2.9 | 0.99 |
| | Machinelike/humanlike | 2.6 | 1.17 |
| | Unconscious/Conscious | 3.4 | 1.34 |
| | Artificial/Lifelike | 2.6 | 1.17 |
| | Moving rigidly/Moving elegantly | 3.6 | 1.42 |
| Animacy ($M$=3.44, $SD$=0.24, Cronbach's Alpha = 0.38) | Dead/Alive | 3.7 | 0.82 |
| | Stagnant/Lively | 3.8 | 0.78 |
| | Mechanical/Organic | 2.9 | 0.99 |
| | Inert/Interactive | 3.1 | 1.37 |
| | Apathetic/Responsive | 3.7 | 1.16 |
| Likeability ($M$=3.88, $SD$=0.29, Cronbach's Alpha = 0.80) | Dislike/Like | 3.2 | 1.03 |
| | Unfriendly/Friendly | 4 | 0.94 |
| | Unkind/Kind | 4.4 | 0.52 |
| | Unpleasant/Pleasant | 3.8 | 1.23 |
| | Awful/Nice | 4 | 0.67 |

| Perceived Intelligence (*M*=3.56, *SD*=0.34, Cronbach's Alpha = 0.73) | Incompetent/Competent | 3.2 | 1.14 |
|---|---|---|---|
| | Ignorant/Knowledgeable | 4 | 0.47 |
| | Irresponsible/Responsible | 3.5 | 0.53 |
| | Unintelligent/Intelligent | 3.3 | 0.82 |
| | Foolish/Sensible | 3.8 | 0.92 |

The overall Cronbach's Alpha score for the virtual human avatar for Godspeed Attributes is 0.83, and with the alpha score being above the threshold of 0.7, sufficient internal consistency and reliability is achieved.

The results from the Godspeed Questionnaire provide a comparative analysis of Nao, the humanoid robot, and Jack, the virtual human avatar, across several attributes: Anthropomorphism, Animacy, Likeability, and Perceived Intelligence. In previous research, Stroessner (2020) reviewed the Godspeed Questionnaire and found the following Cronbach's alpha values for anthropomorphism (0.86-0.93), animacy (0.70-0.76), likeability (0.84-0.92), and perceived intelligence (0.75-0.92). These ranges are also referenced in this study to measure the internal consistency of the results found for each attribute category.

For anthropomorphism, Nao received a mean score of 3.22 (SD = 0.48, Cronbach's Alpha = 0.61), indicating a moderate perception of Nao as natural and humanlike. Individual item scores varied, with "Fake/Natural" and "Unconscious/Conscious" scoring higher, suggesting participants see Nao as somewhat more natural and conscious. However, Cronbach's Alpha of 0.61 suggests questionable internal consistency, indicating that the items may not be perfectly aligned in measuring the same construct.

Jack scored a mean of 3.02 (SD = 0.46, Cronbach's Alpha = 0.70) for anthropomorphism, also indicating a moderate perception of human-likeness but with slightly more consistent responses. The items "Unconscious/Conscious" and "Moving Rigidly/Moving Elegantly" had higher variability, but overall, the internal consistency was acceptable. Both Nao and Jack score similarly on anthropomorphism, indicating moderate perceptions of human-likeness, with Jack's slightly higher Cronbach's Alpha suggesting more consistent responses among the items compared to Nao.

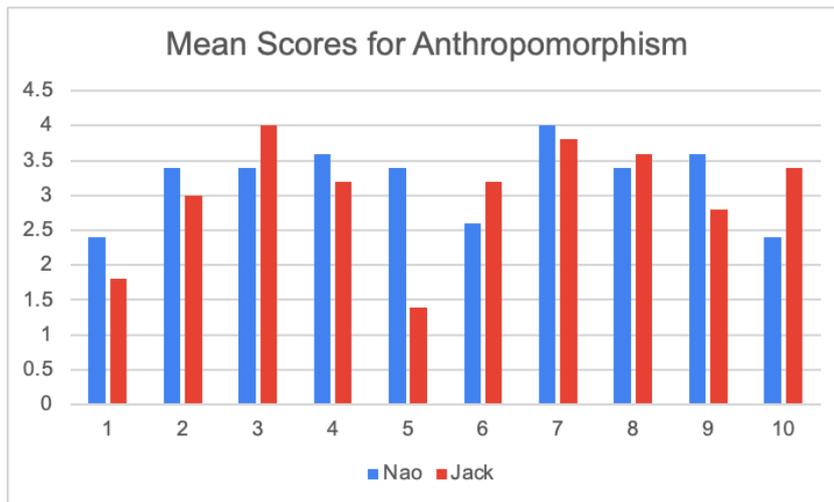

*Figure 4.* *Mean score of each participant for Anthropomorphism dimension*

The animacy dimension shows that Nao received a mean score of 3.9 (SD = 0.71, Cronbach's Alpha = 0.30), reflecting high variability in perceptions of Nao's liveliness, but with poor internal consistency. In comparison, Jack received a mean score of 3.44 (SD = 0.24, Cronbach's Alpha = 0.38), also indicating moderate perceptions of liveliness but with similarly poor internal consistency. Nao scores higher on animacy compared to Jack, suggesting that participants perceive Nao as more lively and interactive, although both assistants have garnered a mixed reception in terms of animacy.

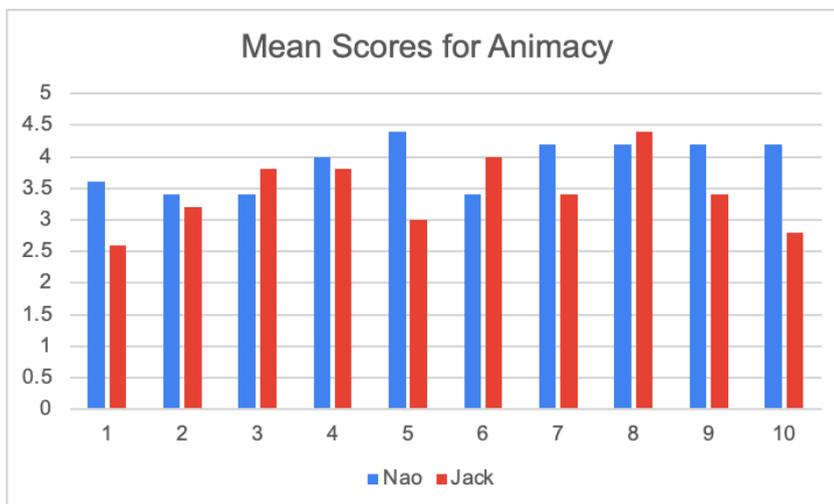

*Figure 5.* *Mean score of each participant for Animacy dimension*

For likeability, Nao achieved a mean score of 4.74 (SD = 0.23, Cronbach's Alpha = 0.76), indicating that participants generally find Nao to be likeable, friendly, and kind, with good internal consistency. Jack, on the other hand, received a mean score of 3.88 (SD = 0.29,

Cronbach's Alpha = 0.80), also showing somewhat positive perceptions but with lower scores than Nao. Both Nao and Jack exhibit acceptable to good internal consistency in likeability, with Nao being perceived as more likeable overall.

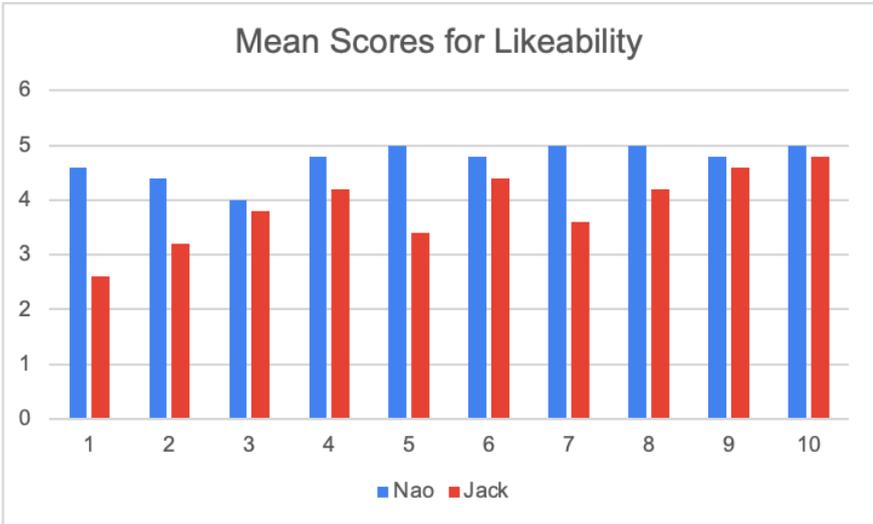

*Figure 6. Mean score of each participant for Likeability dimension*

In the dimension of perceived intelligence, Nao received a mean score of 4.06 (SD = 0.17, Cronbach's Alpha = 0.88), indicating high perceptions of intelligence with excellent internal consistency. Jack received a mean score of 3.56 (SD = 0.34, Cronbach's Alpha = 0.73), suggesting moderate perceptions of intelligence with acceptable internal consistency. Overall, Nao is perceived as more intelligent compared to Jack, with higher mean scores and better internal consistency in this dimension.

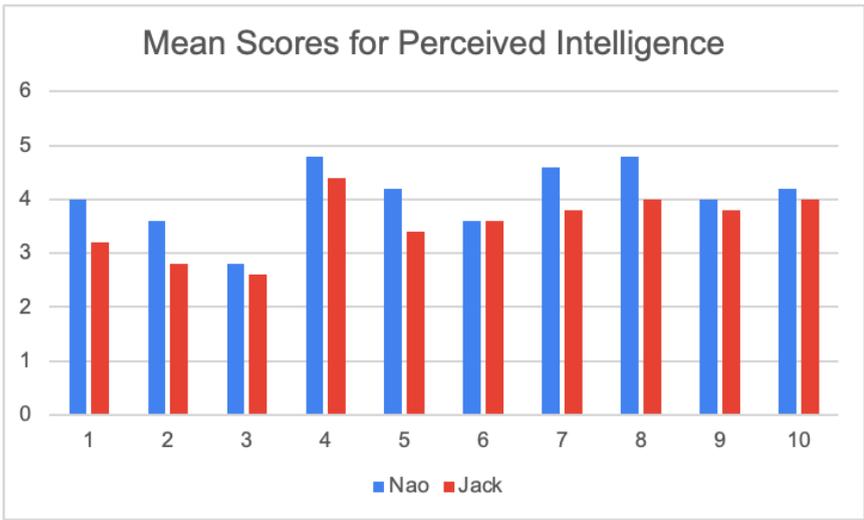

*Figure 7. Mean score of each participant for Perceived Intelligence dimension*

Further inferential statistical analysis was conducted to determine how the humanoid robot Nao and virtual human avatar Jack differ in Godspeed attributes of Anthropomorphism, Animacy, Likeability, and Perceived Intelligence through t-test with the null hypothesis as the difference of the mean scores for each attribute for the two assistants being zero. The aim with this null hypothesis was to determine whether the Godspeed scores for the assistants differ significantly from each other or not. The t-test was performed through Excel Data Analysis tools.

For Anthropomorphism, the null hypothesis failed to be rejected under 0.05 significance level with the p-value of 0.53. This suggests that no significant difference was found between the Anthropomorphism of the humanoid robot Nao, and virtual human avatar Jack in this study. For Animacy, the null hypothesis failed to be rejected under 0.05 significance level with the p-value of 0.05022. However, the null hypothesis can be rejected under 0.1 significance level with 0.9 confidence interval, suggesting that there might be a relation between the perceived Animacy of the two assistants that needs to be examined further. In terms of Likeability, the null hypothesis is rejected under 0.05 significance level with the p-value of 0.002, indicating that the likeability of the two assistants significantly differ from each other, the humanoid robot being liked more with a mean of 4.74 out of 5 score in the Likeability scale. Lastly, for Perceived Intelligence, the null hypothesis fails to be rejected under 0.05 significance level with the p-value of 0.07. However, like Animacy, for this attribute the null hypothesis can be rejected under 0.1 significance level with 0.9 confidence interval, suggesting a moderate difference between the Perceived Intelligence of the two assistants.

To analyze the qualitative data collected from participant interviews, a systematic coding process was employed. Coding involves assigning labels to segments of text that represent different themes or concepts, allowing to identify patterns and draw meaningful insights from the data (Rosala, 2022). The coding process began by carefully reviewing the interview responses and identifying recurring themes related to advice preference, human likeness perception, and comfort levels.

In the coding process for advice preference, several key codes were identified: specific recommendations, general recommendations, natural responses, and context understanding. Participants frequently preferred Nao's advice because it provided specific and tailored recommendations, which felt more natural and sincere compared to Jack's generic responses.

This finding highlights the importance of personalized advice in enhancing the perceived effectiveness of robotic assistants.

For human-likeness perception, codes such as appearance, movements/gestures, and physical presence were identified. Participants noted that while Jack had a more human-like appearance due to his realistic facial features, Nao's natural movements and gestures, along with its physical presence, made interactions feel more genuine. Nao's embodiment in a physical form contributed significantly to its perceived human-likeness, suggesting that interactive behaviors and physical presence are crucial in creating lifelike robotic interactions.

When examining comfort levels, codes such as weird/uneasy feeling, screen interaction, and mechanical voice were used. Some participants reported feeling a sense of weirdness or unease due to the non-human nature of the assistants' eyes and Nao's somewhat mechanical voice. Jack's presence on a computer screen often made interactions feel less personal and more distant, contributing to discomfort. Conversely, Nao's physical presence and interactive gestures provided a more comfortable and engaging experience, mitigating some of the unease associated with interacting with robotic assistants.

Concerning the usability of the systems employed in this study, satisfaction is the most discernible quality of usability that was measured in the scope of the experiment conducted, as defined in ISO 9241-11. The measured scales in the Likeability Godspeed Questionnaire, such as "Unpleasant/Pleasant" and "Dislike/Like" seem directly associated with the formal description of satisfaction which mentions the positive attitudes towards the use of the system at hand (International Organization for Standardization, 1998). Nao has better mean scores throughout the likeability scales with a statistically meaningful margin in comparison to Jack, indicating a higher level of satisfaction with the interactions had with Nao. The other two pillars of usability, effectiveness and efficiency, have been kept relatively consistent throughout the interactions of users with the two assistants Nao and Jack. Considering that effectiveness and efficiency refer to the accuracy and completeness of the achieved goals by the user and the resources consumed in the interaction with the system respectively, the fact that both assistants make use of ChatGPT 3 as their shared model to generate responses and near identical delivery of these responses ensures that these usability metrics are theoretically consistent in the two conditions of the study (International Organization for Standardization, 1998).

The study analyzed the consistency and reliability of data collected via the Godspeed Questionnaire for a humanoid robot (Nao) and a virtual human avatar (Jack) through Cronbach's Alpha scores. Nao and Jack were evaluated on attributes of Anthropomorphism, Animacy, Likeability, and Perceived Intelligence. Nao scored higher in Likeability and Perceived Intelligence, indicating better overall user perception and reliability. Jack showed slightly better internal consistency in Anthropomorphism but both scored similarly. Nao had higher variability and lower internal consistency in Animacy compared to Jack. T-tests indicated no significant difference in Anthropomorphism but significant differences in Likeability and moderate differences in Animacy and Perceived Intelligence favoring Nao. Qualitative analysis revealed preferences for Nao's specific recommendations and natural interactions, highlighting the importance of physical presence and personalized advice in enhancing user experience. Overall, Nao was perceived as more likeable, intelligent, and engaging compared to Jack.

**Discussion and Conclusion**

The results from this study, comparing the humanoid robot Nao and the virtual human avatar Jack on the Godspeed attributes of Anthropomorphism, Animacy, Likeability, and Perceived Intelligence, offer insightful findings that align with existing literature in the field of human-robot interaction. The results indicated no significant difference in Anthropomorphism between Nao and Jack, suggesting that participants perceived both assistants as equally humanlike in their appearance and behaviors. This finding is consistent with previous studies that suggest visual human-likeness can be effectively achieved in both physical and virtual agents, provided that their design and behaviors are sufficiently sophisticated (Hoorn & Huang, 2024). It was previously found that perception of anthropomorphism is more correlated with internal qualities of the robot, rather than the appearance of the robot (Hoorn & Huang, 2024). This supports the results of this study as both agents were designed to have the same internal system to communicate, using the same generative AI model. However, it is important to note that in addition to these inferential results, some participants noted in the interviews that while Jack had a more human-like appearance due to his realistic facial features, Nao's natural movements and gestures, along with its physical presence felt somewhat more human-like. These findings highlight that the embodiment factor can play a role in the perception of human-likeness of an agent, along with the way the agent communicates and the role of embodiment in anthropomorphism may be further investigated in different ways.

For Animacy, the inferential results from the t-test were on the cusp of significance, with a p-value of 0.05022. While the null hypothesis could not be rejected at the 0.05 level, it could be at the 0.1 level, implying a nuanced difference that merits further investigation. This is reflective of the broader literature, which indicates that the perception of good animacy in agents is a complex interplay of movement, responsiveness, and interactive behaviors (Hauser et. al., 2023). The difference found in the means for Animacy suggests that participants might have subtly favored the dynamism of one assistant over the other, potentially influenced by the physical presence and interactive capabilities of Nao compared to the screen-based presence of Jack.

The most notable finding was in the Likeability attribute, where Nao significantly outperformed Jack. This substantial difference suggests that the humanoid robot's physical presence, interactive gestures, and perhaps a more engaging demeanor contributed to a higher likeability score. This could be explained by the differences in appearance of the two assistants, as the appearance of robots were shown to be an critical factor in the warmth associated with the interacted agent (Keskin et al., 2024). The humanoid robot Nao scored high on Likeability scale of the Godspeed Questionnaire in previous studies as well, again suggesting that the appearance of the robot is a significant factor in its high likeability (Keizer et al., 2014). Moreover, the difference in likeability between humanoid robots and virtual humans are also supported by studies such as Rahman (2024), which found that physical embodiment enhances trust and likability in human-robot interactions. The physicality of Nao likely provided a more tangible sense of companionship and social presence, fulfilling basic human social needs more effectively than the virtual avatar. These results can also be understood through the lens of Norman's principles of human-computer interaction. Specifically, Norman's concept of affordances may explain why the physical presence of the humanoid robot led to higher likeability. The robot's physical form provides clear affordances that signal its capabilities and potential actions, making interactions more intuitive and satisfying for users (Norman, 2013).

Perceived Intelligence results showed a p-value of 0.07, indicating a moderate difference that becomes significant at the 0.1 level. This suggests that while both assistants were viewed as intelligent, there might be a slight edge in how one assistant's intelligence was perceived over the other. This aligns with research indicating that perceived competence in robots and virtual agents is influenced not only by their problem-solving abilities and responses but also by their interactive and communicative competencies (Mutlu, 2011). The slight preference could be attributed to the manner in which Nao delivered its advice, appearing more thoughtful and contextually aware, enhancing its perceived intelligence.

These results offer significant implications for the original research questions concerning the comparative effectiveness and user perceptions of humanoid robots versus virtual human avatars. The preference for Nao's advice and higher likeability underscores the importance of physical presence and natural interactive behaviors in human-robot interaction. The nuanced findings in Animacy and Perceived Intelligence suggest that while virtual avatars like Jack can achieve a high level of engagement, physical embodiment still holds a distinct advantage in creating more favorable user experiences. Additionally, the feedback provided by the humanoid robot's physical gestures and expressions likely contributed to its higher perceived intelligence and likeability. Norman (2013) emphasizes the importance of feedback in user interactions, as it helps users understand the outcomes of their actions and feel more connected to the system. The humanoid robot's ability to provide immediate, tangible feedback through its physical presence may have enhanced users' perceptions of its capabilities.

Among the observed results of the study, the comparatively higher level of satisfaction people had in their interaction with Nao is one of the more straightforward ones. In multiple instances participants commented on how cute they found Nao. As mentioned before, Nao outshined Jack in a significant way in the Likeability scale in the Godspeed Questionnaire Series. This result can also be attributed to its relation to the uncanny valley effect. The theory suggests that after a certain point, an increasing amount of humanlike appearance causes a drop in how well liked a robot is. Considering that Jack, though a virtual avatar, still looks more like a human than the humanoid robot Nao, might be perceived less favorably due to the theorized bad reception associated with the uncanny valley effect. Importantly, the discomfort associated with Jack's mere computer screen presence was mentioned by many as well. Even though these are possible contributors to the participants affinity of the two assistants, it is doubtful whether this is the main reason why. At this point, the results for the Godspeed scale Anthropomorphism has to be considered as well. The t-test result indicated that the higher mean score of Jack in this attribute was not sufficient to be deemed significant, meaning that it is likely that other confounding factors are in play that make Nao more liked by participants. Therefore, this also provides an answer to whether or not the uncanny valley effect can be observed by the result of this study. Findings within the scope of this study has proven insufficient to claim the uncanny valley effect at work in this experimental setup.

It is also important to note however, that the perceived effectiveness of the two assistants vary in practice, with some participants reporting the difference in the quality of the responses given by Nao and Jack to similar questions. This does not fit with the expectation of an equal perception of the effectiveness from the participants. Nevertheless, it was observed that even when the controlled variable of an experiment is the intelligence of different robots, the perceived intelligence of these robots may vary, possible explanations being the difference in the robots' animacy among other variabilities (Bartneck et al., 2009).

Overall, the coding and analysis of the data underscored the importance of personalized, contextually appropriate advice, humanlike gestures, and physical presence in creating effective and comfortable interactions with robotic assistants. These findings suggest that future improvements in robotic assistant design should focus on enhancing both the content and embodiment of the assistants to improve user experience and satisfaction.

The practical implications of this study offer valuable insights for robotics designers, researchers, and others in developing robotic and virtual human assistants. Participants' perceived anthropomorphism suggests that physical and virtual agents can achieve high perceived human-likeness with sophisticated design and behaviors. This indicates that focusing on cognitive and communicative abilities is crucial for enhancing user perception of anthropomorphism. The study highlights the importance of physical presence and interactive behaviors. Nao's higher likeability compared to Jack, attributed to its physical presence and interactive gestures, indicates that physical robots can provide a more tangible sense of companionship and social presence. UX designers can incorporate physical elements and interactive features to enhance the likeability and social presence of robotic assistants. The perceived intelligence results showed a moderate difference, with Nao's physical gestures and expressions contributing to a higher perceived intelligence. Designers should focus on providing immediate, tangible feedback through physical gestures and expressions to enhance users' perceptions of an agent's capabilities.

It should be mentioned that the relatively small sample size of the study may limit the generalizability of the findings. Additionally, the narrow demographic range may not represent the broader population only consisting of students from the same university, faculty and age range. Future studies could benefit from a larger and more diverse sample to enhance the external validity of the results. While the Godspeed Questionnaire Series is widely used in HRI research, future studies could consider using multiple measures or creating custom scales to better capture the nuances of user perceptions for specific contexts. Authors of the Godspeed Questionnaire Series draw attention to the importance of having participants fill out the questionnaire in their native language, even when participants are proficient in English (Bartneck, 2023). For this reason, a limitation of this study conducted on all Turkish participants is the lack of an official Turkish translation of the questionnaire. The text-to-speech technology used in the two assistants are not the same, Nao uses the Google service while Jack uses OpenAI Whisper. Even though erroneous transcriptions of what was said by the participants happened very infrequently, this distinction in the methodology of the study poses a possible liability. Future research could look into ways of comparing two assistants that both use the same technology as a text-to-speech service. The field of prompt engineering also outlines an aspect of this study where there is room for improvement. As mentioned, prompts are critical in how an LLM behaves, therefore a prompt was created with this in mind (White et al., 2023). However, prompt engineering is a new concept with great potential to help programmers have better control over the general behavior of LLMs and receive outputs to their liking. That is precisely why further researchers should look into the growing area of prompt engineering for better conversational assistants in their studies.